%% file: ms.tex
\documentclass[3p,12pt]{elsarticle}
\usepackage{amssymb} 
\usepackage{pgf}
\usepackage{subcaption}
\usepackage[version=3]{mhchem}
\usepackage{multirow}
\usepackage[nolist]{acronym}
\begin{acronym}
	\acro{AMR}{adaptive mesh refinement}
	\acro{DNS}{direct numerical simulation}
	\acroplural{DNS}[DNS]{direct numerical simulations}
	\acro{1D}{one-dimensional}
	\acro{2D}{two-dimensional}
	\acro{IFI}{intrinsic flame instability}
	\acroplural{IFI}[IFIs]{intrinsic flame instabilities}
	\acro{TDI}{thermodiffusive instability}
	\acroplural{TDI}[TDIs]{thermodiffusive instabilities}
	\acro{H2}[\ce{H2}]{hydrogen}
	\acro{H2O}[\ce{H2O}]{water}
	\acro{NH3}[\ce{NH3}]{ammonia}
	\acro{NO}[\ce{NO}]{nitrogen oxide}
	\acro{NOx}[NO\textsubscript{x}]{nitrogen oxide}
	\acroplural{NOx}[NO\textsubscript{x}]{nitrogen oxides}
	\acro{jPDF}{joint probability density function}
\end{acronym}
\usepackage[capitalise]{cleveref}
\usepackage{longtable}
\usepackage{siunitx}
\usepackage{xr}
\usepackage{subcaption}
\newcommand{\XHF}{X_{\rm H_2,F}}
\newcommand{\lF}{l_{\rm F}}

\newcommand{\rmd}{\mathrm{d}}
\newcommand{\Tu}{T_{\rm u}}

\graphicspath{{..}{Figures}}
\biboptions{sort&compress}

\let\today\relax
\makeatletter
\def\ps@pprintTitle{%
	\let\@oddhead\@empty
	\let\@evenhead\@empty
	\def\@oddfoot{\footnotesize\itshape
		{Submitted preprint} \hfill\today}%
	\let\@evenfoot\@oddfoot
}
\makeatother

\begin{document}
	
\begin{frontmatter}
\title{Effects of Intrinsic Flame Instabilities on Nitrogen Oxide Formation in Laminar Premixed Ammonia/Hydrogen/Air Flames}

\author[fir]{Terence Lehmann\corref{cor1}}

\author[fir]{Nikita Dimidziev}
\author[fir]{Thomas L. Howarth}
\author[fir]{Michael Gauding}
\author[fir]{Heinz Pitsch}

\address[fir]{Institute for Combustion Technology, RWTH Aachen University, Templergraben 64, 52056 Aachen, Germany}
\cortext[cor1]{Corresponding author: t.lehmann@itv.rwth-aachen.de}
\input{chapters/00_Abstract}
\end{frontmatter}

\input{chapters/01_Introduction}
\input{chapters/02_ConfigurationNumerics.tex}
\input{chapters/03_Results}
\input{chapters/04_Conclusions}
\input{chapters/05_Acknowledgements}

\input{ms.bbl}
\clearpage
\renewcommand\thefigure{A.\arabic{figure}}    
\setcounter{figure}{1}
\renewcommand\thetable{A.\arabic{figure}}    
\setcounter{table}{1}
\input{chapters/99_Supplementary}

\end{document}

%% file: chapters/00_Abstract.tex
\begin{abstract} 
	
This study investigates the characteristics of \ac{NO} formation in \ac{2D} laminar premixed \acl{NH3}/\acl{H2}/air flames and the impact of thermo-diffusively driven \acp{IFI}. To this end, a set of three highly resolved \acp{DNS} at lean ambient conditions and varying hydrogen fraction in the fuel blend are conducted. The analysis of these \acp{DNS} reveals a significant increase of \ac{NO} formation in positively curved regions of the flame, particularly for lower hydrogen fuel fractions, while negatively curved areas exhibit reduced NO concentrations. However, despite the strong variations of local mass fractions of \ac{NO} in the flame sheet, the mean mass fraction in the post-flame region remains close to the solution from a \acl{1D} flame. Through a representative flame segment analysis of positively curved, negatively curved, and flat regions, key reactions contributing to NO formation are determined, with the \ce{HNO} pathway being the predominant production and the de\acs{NOx} pathway being the predominant consumption pathway across all cases. Thermal \ac{NO} plays no significant role in the considered cases. Generally, the peaks of \ac{NO} production shift to lower values of progress variable in the negatively curved regions, leading to an annihilation of the production and consumption terms in the low hydrogen fuel fraction case. The decrease of \ac{NO} production is found to be mainly driven by changes of the radical concentrations, rather than changes of the temperature-dependent reaction rate coefficients.
\acresetall	
\end{abstract}

%% file: chapters/01_Introduction.tex
\section{Introduction}
The search for adequate replacements of fossil fuels has attracted substantial research efforts, with \ac{H2} and \ac{NH3} being identified as two promising candidates~\cite{Kobayashi2019}. While pure \ac{H2} is difficult to store and transport, and pure \ac{NH3} presents poor combustion properties such as low burning velocities and narrow flammability limits, blends of \ac{NH3}/\ac{H2}, obtained from partial cracking of \ac{NH3}, are a widely considered option for practical use cases~\cite{Kobayashi2019, Elbaz2022, ValeraMedina2018}. However, despite the absence of \ce{CO2} emissions during combustion of \ce{NH3}/\ce{H2} blends, the formation of \acfp{NOx} for such fuels presents a major hindrance for their wide spread application~\cite{Netzer2021, Karimkashi2023, Awad2023}.

\begin{figure}[h!]
	\centering
	\includegraphics[scale=1]{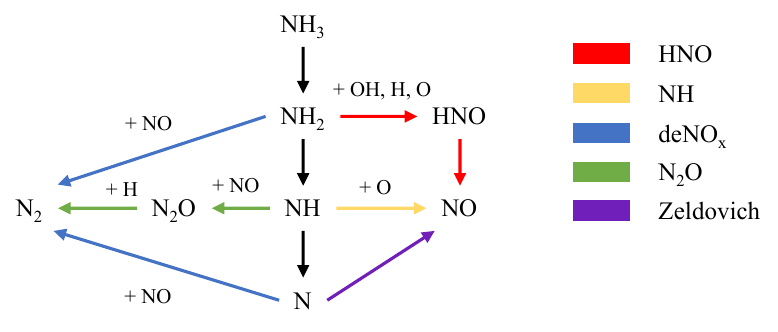}
	\caption{Summary of the most important \ac{NO} production and consumption pathways for \ac{NH3} combustion following \citet{Glarborg2018} and \citet{Zhang2021}. For simplicity, some reactions are lumped into a joint pathway for this representation.}
	\label{fig:NO_mech}
\end{figure}

Numerous studies concerning \ac{NOx} emission characteristics of premixed \ce{NH3}/\ce{H2} flames can be found in the literature. One of the most detailed summaries of nitrogen chemistry was given by \citet{Glarborg2018}. \Cref{fig:NO_mech} depicts a summary of the discussed \ac{NO} production and consumption pathways. Here, \ce{NH2}, produced from \ac{NH3}, reacts via \ce{HNO}, or after a second \ce{H}-abstraction, via \ce{NH} to \ce{NO}. These pathways are referred to as \ce{HNO}, and \ce{NH} pathways, respectively. Together, they describe the formation of fuel-\ac{NO}. In competition with these formation pathways, \ce{NO} can also be consumed via reactions with \ce{NH2} and \ce{NH}. These pathways are referred to as de\acs{NOx} and \ce{N2O} pathways. The Zeldovich mechanism describes the formation of \ac{NO} via the \ce{N} radical, which is either produced from \ce{NH}, or via the so-called thermal pathway \ce{N2 + O = NO + N}.

\citet{Duynslaegher2009} conducted experimental work on planar laminar premixed flames, evaluating flame structure and identifying a profound impact of equivalence ratio $\phi$ on \ac{NO} emissions, particularly pronounced for $\phi \in [0.9, 1]$. In their experimental work, \citet{Li2014} identified fuel \ac{NO} as the dominant contributor to \ac{NO} production over thermal \ac{NO} for stoichiometric and rich flames, considering different fuel blends ($\XHF \in [0.33, 0.6]$) and equivalence ratios ($\phi \in \{1, 1.1, 1.25\}$). The molar fraction of \ac{H2} in the \ac{H2}/\ac{NH3} fuel blend, is defined as 

\begin{equation}
	\XHF=\frac{X_{\ce{H2}}}{X_{\ce{H2}} + X_{\ce{NH3}}}\,.
	\label{eq:XHF}
\end{equation} 

\citet{Nie2023} confirmed the pronounced impact of equivalence ratio on \ce{NO} formation in \ce{NH3}/\ce{H2} flames in their numerical studies for a fuel blend with $\XHF = 0.3$. Utilizing simulations of shock tubes and laminar premixed counterflow flames, \citet{DaRocha2019} compared different reaction mechanisms for the combustion of \ce{NH3}/\ce{H2} blends. The authors carried out sensitivity analyses to identify dominant reaction pathways of \ac{NO} formation. For a fuel blend with $\XHF = 0.2$, abundance of \ce{O} and \ce{OH} radicals, associated with the fuel-\ac{NO} pathway, was found to be the main driver of \ac{NO} emissions. Furthermore, they show that the \ac{NO} concentrations for pure components, i.e., pure \ac{NH3} or pure \ac{H2}, are lower than those of blends.

By carrying out simulations of \ac{1D} premixed flames for different fuel blends, \citet{Zhang2024} employed sensitivity analyses to identify the main contributing reactions for \ce{NO} formation and consumption. The authors established a relation between \ce{NO} formation levels and maximum observable \ce{OH} concentrations, in line with experimental findings by \citet{Shi2022}. Higher peak concentrations of \ce{OH} were linearly correlated with \ce{NO} formation for equivalence ratios $\phi > 1$. Sensitivity analyses as well as reaction pathway analyses for $\phi = 1.1$ and $\XHF \in \{0.1, 0.3, 0.5\}$ identified the \ce{HNO} path associated with fuel-\ce{NO} as dominant for \ac{NO} formation. \Ac{NO} consumption was strongly driven by reactions involving amine radicals \ce{NH2} and \ce{NH}. Furthermore, the importance of the \ce{NNH} radical for \ce{NO} consumption was underlined and \ce{NO} emissions were shown to increase with increasing $\XHF$.

Several previous studies have demonstrated that \ce{NH3}/\ce{H2} blends are susceptible to \acp{IFI}~\cite{Kobayashi2019, Lehmann2024}. \citet{Netzer2021} studied the impact of artificially induced flame wrinkling on \ac{NO} production in premixed \ce{NH3}/\ce{H2} flames for different equivalence ratios. In agreement with the research mentioned previously, a high spatial correlation between concentrations of \ce{OH} and \ac{NO} was found. Local first order correlations showed the \ac{NO} concentration to be correlated with oxygen mass fraction $Y_{\ce{O2}}$, local equivalence ratio, and the sign of the local curvature. The high diffusivity of \ce{H2} and its consequent impact on the local radical pool in cellular structures caused by flame instabilities was identified as a crucial factor for local \ce{NO} formation.

\citet{Karimkashi2023} investigated the \ac{NOx} formation in premixed \ce{NH3}/\ce{H2} flames, considering \ac{1D} laminar flames for different fuel blends of $\XHF \in \{0, 0.4, 0.6\}$ as well as one 3D turbulent simulation at $\XHF = 0.4$ and stoichiometric equivalence ratio. The authors defined different pathways for \ac{NO} production consisting of sets of elementary reactions and quantified their impact on the net \ac{NO} formation. For all considered fuel blends in the \ac{1D} case, the authors identified the \ce{HNO} pathway to be the dominant production, and the \ce{N2O} pathway to be the dominant consumption route. The impact on the radical pool of \ce{H2} addition was found to significantly affect \ce{NO} formation, with higher concentrations of \ce{H2} leading to more radical formation and as such a promotion of \ce{NO} formation through the \ce{HNO} pathway. Investigations of turbulent flames led the authors to the conclusion that preferential diffusion of \ce{H2} plays a major role in local characteristics of \ce{NO} formation. \citet{Rieth2023} analyzed \ac{NO} formation in turbulent \ac{NH3}/\ac{H2} flames at different equivalence ratios and \citet{DAlessio2024b} considered a variation of the pressure.

The present work aims to complement existing research of \ac{NO} formation in premixed \ce{NH3}/\ce{H2}/air flames by analyzing the effect of $\XHF$ in laminar \ac{2D} flames under the influence of \acp{IFI}. To this end,  three cases with low, medium, and high $\XHF$ are assessed. This work is structured as follows. First, the numerical methods, models, and the configuration are detailed in the methodology section. Additionally, a method for direct comparison between \ac{1D} and \ac{2D} simulations, hereafter referred to as flame segment analysis, is introduced. Next, results for highly resolved \ac{2D} \acp{DNS} are discussed based on an analysis of global \ac{NOx} characteristics as well as a flame segment analysis considering distinct \ce{NO} formation pathways. The paper closes with the conclusions.

%% file: chapters/02_ConfigurationNumerics.tex
\section{Methodology}

\subsection{Numerical methods and models}

The \ac{DNS} are conducted using PeleLMeX~\cite{PeleLMeX_JOSS, PeleSoftware}. The code solves the multi-species reactive Navier-Stokes equations within a low-Mach limit~\cite{day2000numerical}. The temporal advancement of these equations is achieved through a spectral-deferred correction method that ensures the conservation of species, mass, and energy~\cite{nonaka2012deferred, nonaka2018conservative}. The advection term is discretized using a second-order Godunov scheme. Energy and species equations are treated implicitly with the ODE solver CVODE from the SUNDIALS package~\cite{hindmarsh2005sundials}. Additionally, PeleLMeX incorporates \ac{AMR} capabilities inherited from the AMReX package~\cite{AMReX_JOSS}, which is used to locally refine the flame front as described below.

Chemical reactions and their associated rates are modeled using the reaction mechanism developed by Zhang et al.~\cite{Zhang2021}, which includes 30 species and 243 reactions. This model has shown strong overall performance, demonstrating good quantitative agreement with extensive experimental data regarding flame speed, ignition delay time, and species concentrations~\cite{Girhe2024}. To avoid assumptions related to chemical mechanism reduction, detailed chemistry is utilized. Transport properties are modeled using a mixture-averaged approach including the Soret effect~\cite{Howarth2024}. For details, the reader is referred to~\citet{Lehmann2024}.

\subsection{Configuration}

\ac{2D} \ac{DNS} are carried out in a rectangular domain with dimensions of $L_y=400\lF$ in streamwise ($y$) and $L_x=200\lF$ in crosswise ($x$) direction. Inflow and outflow conditions are applied at $y=0$ and $y=400\lF$, respectively, while a periodic condition is applied in $x$ direction. Here, $\lF$ is the thermal flame thickness defined as $\lF = \left(T_{\rm b} - T_{\rm u}\right)/\left(\max\left(\rmd T / \rmd x\right)\right)$ in an unstretched premix flame where $T$ denotes the temperature and the indices $\rm u$ and $\rm b$ denote the values in the unburned and burned, respectively. The grid has a base resolution of $n_x\times n_y = 768 \times 1536$ with three levels of \ac{AMR} leading to a fine resolution of $\Delta x_{\rm Fine} \approx \lF / 30$ within the flame zone. To initialize the simulation, a \ac{1D} flamelet computed with FlameMaster~\cite{Pitsch1998} is mapped onto the domain and perturbed by a series of $N=200$ harmonic functions~\cite{AlKassar2024}, so that the position of the flame is given by
\begin{equation}
	y_{\rm Flame}(x, t=0) = y_0 + A_0 \sum_{i=1}^{N} \sin\left(i \frac{2\pi}{L_x} x + \psi(i) \right)\,.
\end{equation}
Here, $y_0=360\lF$ is the initial location of the mean flame front, $A_0=0.01\lF$ is the amplitude of each frequency, and $\psi(i)\in\left[0, 2\pi\right)$ is a random phase shift obeying a uniform distribution. This initialization equally excites all possibly relevant scales of \acp{IFI} and leads to a fast transition to a fully developed statistically steady flame.

Within the scope of this study, three conditions are examined. For all cases, the equivalence ratio is $\phi=0.6$ at ambient pressure ($p=\SI{1}{\bar}$) and an inlet temperature of $\Tu=\SI{298}{\kelvin}$. The molar fraction of \ac{H2} in the fuel blend, as in \cref{eq:XHF} is chosen as $\XHF\in\{0.4, 0.6, 0.8\}$. These cases will be referred to as low, medium, and high $\XHF$ cases hereafter.

\subsection{Flame segment construction}

To facilitate the comparison between an unstretched \ac{1D} reference flame and the local flame state in the \ac{2D}, \ac{1D} segments are computed from the \ac{2D} simulation following the approach of \citet{Day2009} and \citet{Wen2023}. In contrast to a direct evaluation of the reaction rates on the full domain, the construction of segments allows for a correlation of the pre- and post-flame zone with the local curvature at the flame front. The starting points of the segments are chosen along the isoline of the progress variable based on \ac{NH3}, $C_{\ce{NH3}} = 0.95$. The segments are then constructed along the positive and negative gradient of the progress variable $C_{\ce{H2O}}$ based on \ac{H2O} using a fourth-order Runge-Kutta method and quadratic interpolation with a step size of $\Delta s = 0.1\Delta x_{\rm Fine}$. The segments are terminated at $C_{\ce{H2O}}<1\cdot 10^{-2}$ at the lower end, while a minimal gradient of the progress variable criterion of $\Delta C_{\ce{H2O}} / \Delta s \leq 1\cdot 10^{-3}$ is imposed for termination at the upper end. Finally, the species mass fractions are quadratically interpolated onto the segments. Both $C_{\ce{NH3}}$ and $C_{\ce{H2O}}$ are defined as $C_i = \left(Y_i - Y_{i, \rm u}\right) / \left(Y_{i, \rm b} - Y_{i, \rm u}\right)$. $C_{\ce{NH3}} = 0.95$ yields an accurate representation of the peak heat release location and with this the flame geometry, and hence is a suitable choice for the segment's starting points. However, it cannot resolve the \ac{NO} production zone as \ac{NH3} is fully consumed ($C_{\ce{NH3}} = 1$) before the source term of \ac{NO} approaches zero. On the other hand, $C_{\ce{H2O}}$ is unsuitable to represent the flame geometry but exhibits non-zero gradients across the whole \ac{NO} production zone.

%% file: chapters/03_Results.tex
\section{Results and discussion}

Within this section, the \ac{2D} laminar premixed flames are analyzed with respect to \ac{NO} emissions and compared to a \ac{1D} unstretched premixed flame, hereafter referred to as \ac{1D} flame. First, the global \ac{NO} formation characteristics are examined, whereafter a flame segment analysis is conducted.

\subsection{\ac{NO} formation characteristics}

\begin{figure}[h!]
	\centering
	\includegraphics[scale=1]{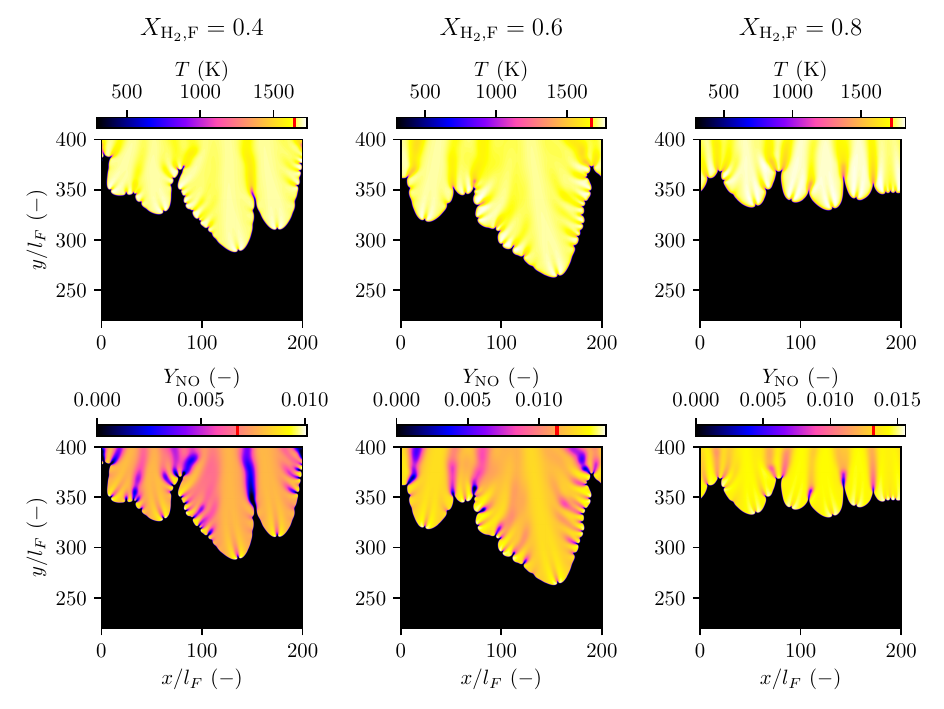}
	\caption{Fields of temperature $T$ (top row) and NO mass fraction $Y_{\ce{NO}}$ (bottom row) for hydrogen fractions of $\XHF= 0.4, 0.6, 0.8$ (left to right). Red lines on the color bar represent the burned state in a \ac{1D} unstretched premixed flame at the same conditions. Note that the scales are different across the cases to capture their individual details. Conditions: $\phi=0.6, \Tu=\SI{298}{\kelvin}, p=\SI{1}{\bar}$.}
	\label{fig:fields}
\end{figure}

\Cref{fig:fields} shows the distribution of temperature $T$ and \ce{NO} mass fraction $Y_{\ce{NO}}$ for $\XHF= 0.4, 0.6,$ and $0.8$. From the temperature fields, it becomes evident that in all cases the flame is experiencing \acp{TDI}, showing the characteristic flame fingers described in other works~\cite{Berger2019, Howarth2022}. The positively curved regions, i.e., regions where the flame front is convex towards the unburned, show super-adiabatic temperatures with a relative increased of $5.2\%$, $6.0\%$, and $5.9\%$ for the low, medium, and high $\XHF$ cases, respectively, compared to the \ac{1D} flame. In the negatively curved regions, i.e., regions where the flame front is concave towards the unburned, the temperature falls below the adiabatic flame temperature. The low $\XHF$ case appears to show the smallest scales of flame structures, which increase with increasing $\XHF$.

The mass fractions of \ce{NO} in \cref{fig:fields} generally follow the shape of the flame fingers. Across all cases, $Y_{\ce{NO}}$ is increased in the positively curved regions compared to the \ac{1D} flame. This effect is the strongest for the low $\XHF$ case with a relative increase of 49\%, compared to 31\% and 18\% in the medium and high $\XHF$ cases, respectively. At the same time, the high $\XHF$ case shows the highest absolute values of $Y_{\ce{NO}}$ in both the \ac{1D} and \ac{2D} simulation. In the negatively curved regions, $Y_{\ce{NO}}$ appears to be significantly reduced, showing long trails into the burned region. It is worth mentioning that mass fractions of \ce{NO2}, often considered together with \ac{NO} as \ac{NOx}, is several orders of magnitude lower that that of \ac{NO}.

\begin{figure}[h!]
	\centering
	\includegraphics[scale=1]{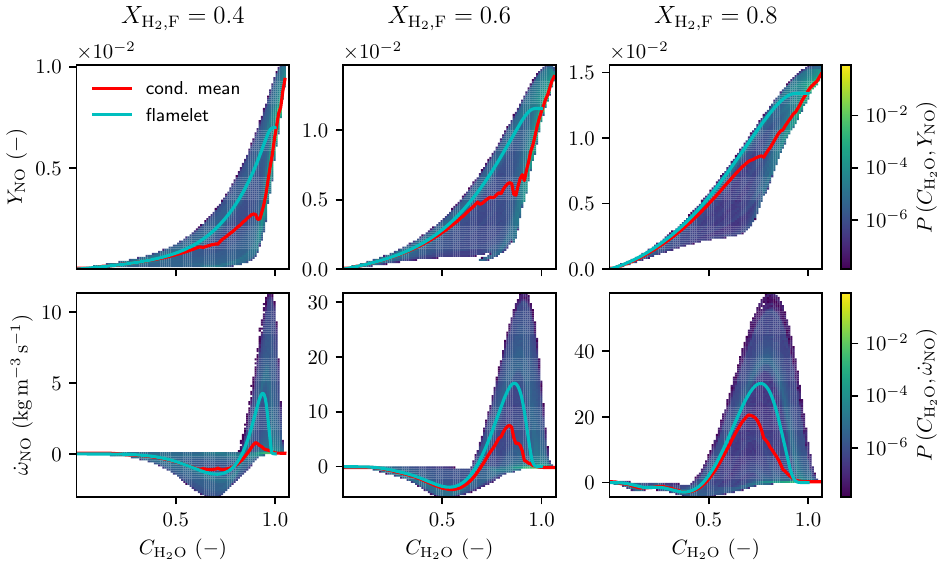}
	\caption{\Acfp{jPDF} between $C_{\ce{H2O}}$ and $Y_{\ce{NO}}$ (top row), $C_{\ce{H2O}}$ and the source term of \ce{NO}, $\dot{\omega}_{\ce{NO}}$ (bottom row). Red and cyan lines represent the conditional mean and an unstretched \ac{1D} profile, respectively. Note that the scales are different across the cases to capture their individual details.}
	\label{fig:statistics}
\end{figure}

\cref{fig:statistics} shows the \acfp{jPDF} of $Y_{\ce{NO}}$ and $\dot{\omega}_{\ce{NO}}$ with $C_{\ce{H2O}}$. The \ac{jPDF} between $C_{\ce{H2O}}$ and $Y_{\ce{NO}}$ (top row) shows a considerable spread for the low $\XHF$ case whereas this spread is reduced for the high $\XHF$ case. More specifically, the \ac{jPDF} reveals that for low $\XHF$, regions with $Y_{\ce{NO}}\approx 0$ exist for up to $C_{\ce{H2O}}=0.8$, whereas this is not the case for the high $\XHF$ case. A similar behavior is observed for the source term of \ac{NO}, $\dot{\omega}_{\ce{NO}}$ (bottom row), where the relative difference between the \ac{1D} flame and the peak source term in the \ac{2D} flame decreases with increasing $\XHF$. Furthermore, the flamelet as well as the \ac{2D} simulation show a region of \ac{NO} consumption at low progress variable followed by a narrow production zone in the low $\XHF$ case. This can be explained by the produced \ac{NO} diffusing upstream, where it is consumed. With increasing $\XHF$, the production zone is broadened and the consumption zone is flattened. 

\begin{figure}[h]
	\centering
	\includegraphics[scale=1]{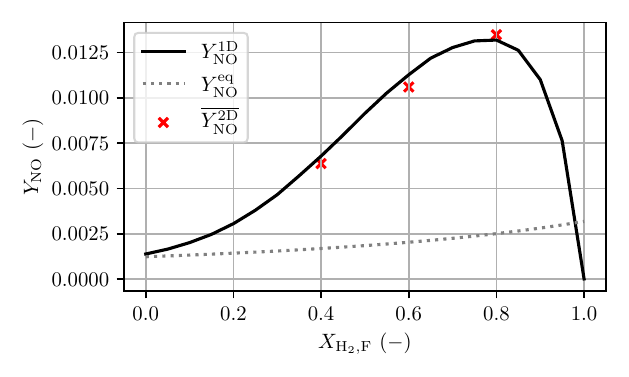}
	\caption{Mass fractions of \ac{NO} in the equilibrium ($Y_{\rm NO}^{\rm eq}$), the burned region of a \ac{1D} flame ($Y_{\rm NO}^{\rm 1D}$), and in the post flame region of a \ac{2D} flame as defined in \cref{eq:mean_no} ($\overline{Y_{\rm NO}^{\rm 2D}}$) over $\XHF$.}
	\label{fig:1DComparison}
\end{figure}

It is important to notice that for all three cases, the conditional mean of $Y_{\ce{NO}}$ is lower than the flamelet solution for $C_{\ce{H2O}}<1$. Similarly, the conditional mean of $\dot{\omega}_{\ce{NO}}$ in the \ac{2D} flame shows lower production rates compared the flamelet, especially for the low $\XHF$ case. Since the values of $C_{\ce{H2O}}$ do not distinguish the unburnt from the burnt gas region, i.e., can be both greater and smaller than unity, the mean mass fraction of \ac{NO} in the post flame region defined as 
\begin{equation}
	\overline{Y_{\rm NO}^{\rm 2D}} = \left.\frac{m_{\ce{NO}}}{m}\right|_{\rm b} = \frac{\int_{\Omega_{\rm b}}\rho Y_{\ce{NO}}\rmd V}{\int_{\Omega_{\rm b}}\rho \rmd V}\,,
	\label{eq:mean_no}
\end{equation}
is considered to draw conclusions for the exhaust gas composition. Here, $m_{\ce{NO}}$ and $m$ are the mass of \ce{NO} and the total gas, respectively, $\rho$ is the density, and $\Omega_{\rm b}$ is the burned region downstream of \ac{NO} production or consumption. However, defining $\Omega_{\rm b}$ is not trivial since $C_{\ce{NH3}}$ approaches zero within the \ac{NO} formation regime and $C_{\ce{H2O}}$ is not bounded. For simplicity, $\Omega_{\rm b}$ is thus defined as the area between $y=395\lF$ and $y=400\lF$, and \cref{eq:mean_no} is averaged over multiple timesteps until it converges to a statistically stationary solution. Since by definition, $\dot{\omega}_{\ce{NO}}$ is negligible anywhere within $\Omega_{\rm b}$, the mean mass is constant and consequently does not depend on the exact formulation of the area. \Cref{fig:1DComparison} shows the results for $\overline{Y_{\rm NO}^{\rm 2D}}$ along with the exhaust value of \ac{1D} flames, $Y_{\rm NO}^{\rm 1D}$, and the equilibrium \ac{NO} mass fraction $Y_{\rm NO}^{\rm eq}$ over $\XHF$. Despite the strong increase of local \ac{NO} mass fractions in the low $\XHF$ case (\cref{fig:fields}), $\overline{Y_{\rm NO}^{\rm 2D}}$ is 5\% smaller than the \ac{1D} value, while only an increase of 5\% is observed for the high $\XHF$ case. This effect can be explained by the lower temperature and hence higher density in the negatively curved regions, and consequently their strong impact on $\overline{Y_{\rm NO}^{\rm 2D}}$. Since $Y_{\rm NO}$ in the negatively curved regions is strongly decreased, this reduces $\overline{Y_{\rm NO}^{\rm 2D}}$. To further examine the origin in the difference between \ac{NO} formation in positively and negatively curved regions, a flame segment analysis is carried ot in the following section. Finally, it is worth mentioning that $Y_{\rm NO}^{\rm eq} < Y_{\rm NO}^{\rm 1D}$ for most blends. This is related to the partial equilibrium of the fuel-\ac{NO} pathways results in \ac{NO} concentrations higher than the global equilibrium. The subsequent decrease to $Y_{\rm NO}^{\rm eq}$ is very low, so that no significant gradients are observed. 

\subsection{Flame segment analysis}

\Acl{1D} flame segments are constructed as outlined in the methodology section, resulting in 22911 segments across the flame for a single timestep. These segments fully cover the flame area relevant to \ac{NO} consumption and production (see \cref{fig:selected_segments} in the supplementary material). To assess the impact of local curvature at the flame front on \ac{NO} production, the segments are then categorized into positive, negative, and neutral curvature segments based on their individual curvature at $C_{\ce{NH3}}=0.95$. Here, the curvature is defined as divergence of the normal vector of $C_{\ce{NH3}}$ which points towards the unburned mixture,
\begin{equation}
	\kappa = -\nabla \cdot \frac{\nabla C_{\ce{NH3}}}{\left|\nabla C_{\ce{NH3}}\right|}\,.
\end{equation} 
The threshold is defined as $\kappa \geq \max(\kappa)\cdot 10^{-2}$ for positively curved segments and $\kappa \leq \min(\kappa)\cdot 10^{-2}$ for negatively curved segments. From each of the defined categories, one representative segment is chosen for the following analysis. An assessment of the statistical validity of this procedure is depicted in \cref{fig:segment_stats} in the supplementary material. As expected from the preceding analysis, the conditional mean  $\langle \dot{\omega}_{\ce{NO}} | C_{\ce{H2O}} \rangle$ for the group of segments in the positively and negatively curved regions show a significant difference. While the conditional mean of the segments in positive curvature regions is relatively close to the flamelet solution, they strongly deviate in the negative curvature region. In both cases, the selected representative flame segment lies within one standard deviation from the mean of that curvature range.

Based on the selected segments, the \ac{NO} formation pathways are analyzed while accounting for the dependence on local curvature. In a first step, the relevant reactions for \ac{NO} formation are selected from the mechanism with a threshold of 5\% contribution to the integrated net NO production or consumption. Following the approach of other studies~\cite{Karimkashi2023, Rieth2023, DAlessio2024b}, these reactions are then grouped into the five pathways introduced in \cref{fig:NO_mech}, namely the \ce{HNO}, \ce{NH}, de\ce{NO_$x$}, \ce{N2O}, and Zeldovich pathways. A list of the considered reactions and their associated pathway is given in \cref{supp_tab:reactions} of the supplementary material. \Cref{fig:1DPaths_xh2_040} shows the production rates $\dot{\omega}_{\mathrm{NO}, i}$ for the individual pathways, their sum, and the full mechanism. In all cases, the sum of the pathways, i.e., the production through the subset of selected reactions, aligns well with the production rate calculated based on the full mechanism, indicating that no relevant reactions are neglected. For all cases (flamelet, neutral, positive, and negative curvature), the \ce{HNO} pathway is dominant for production followed by the \ce{NH} pathway. The de\ac{NOx} pathway is the major consumer of \ac{NO} followed by the \ce{N2O} path. The Zeldovich pathway shows only minor negative to no contribution. These observations are in agreement with \citet{Rieth2023}. 
The production rates of the \ac{1D} solution are very similar to the segment from the close-to-zero curvature area. Although the pathways for the segment in the positively curved region show overall higher production rates and are shifted to higher progress variables, their relative importance remains unchanged. For the segment from the negatively curved region, the overall production rates decrease. At the same time, the relative importance of the de\ac{NOx} pathway increases. Furthermore, the peaks of the production and consumption pathways align in the $C_{\ce{H2O}}$ space, leading to a decrease of the net \ac{NO} formation.

\begin{figure}[h!]
	\centering
	\includegraphics[scale=1]{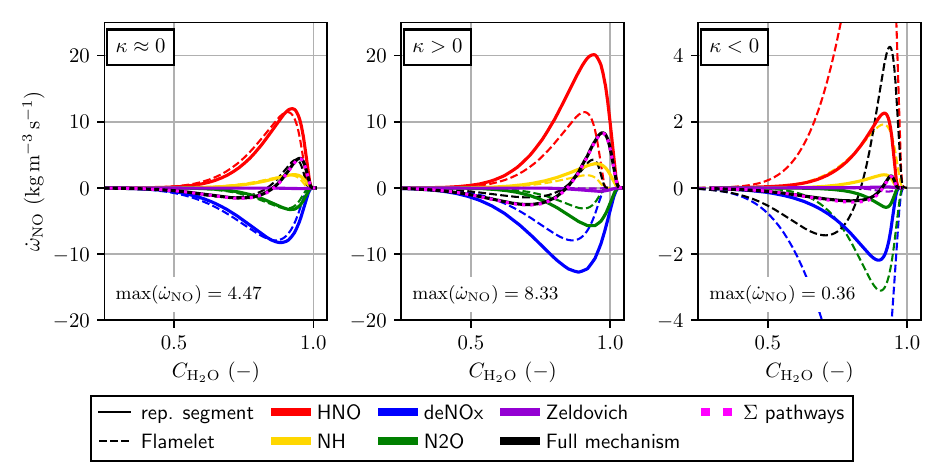}
	\caption{Low $\XHF$ case: Net production rates $\dot{\omega}_{\mathrm{NO}, i}$ for the individual pathways, their sum, and the full mechanism. Top left: Flamelet, top right: neutral curvature, bottom left: positive curvature, bottom right: negative curvature. Note that the $y$-scale for the bottom right figure is adjusted to capture the details of the individual segments. Conditions: $\XHF=0.4, \phi=0.6, \Tu=\SI{298}{\kelvin}, p=\SI{1}{\bar}$.}
	\label{fig:1DPaths_xh2_040}
\end{figure}

\begin{figure}[h!]
	\centering
	\includegraphics[scale=1]{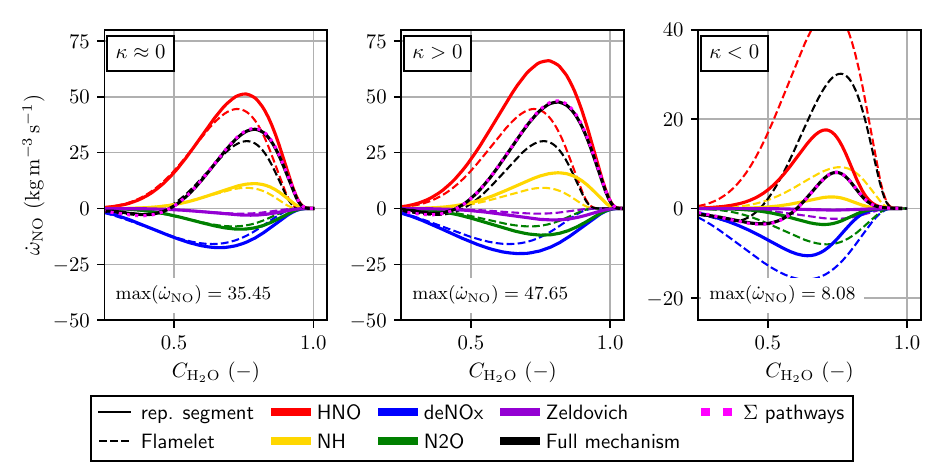}
	\caption{High $\XHF$ case: Net production rates $\dot{\omega}_{\mathrm{NO}, i}$ for the individual pathways, their sum, and the full mechanism. Top left: Flamelet, top right: neutral curvature, bottom left: positive curvature, bottom right: negative curvature. Note that the $y$-scale for the bottom right figure is adjusted to capture the details of the individual segments. Conditions: $\XHF=0.8, \phi=0.6, \Tu=\SI{298}{\kelvin}, p=\SI{1}{\bar}$.}
	\label{fig:1DPaths_xh2_080}
\end{figure}

\Cref{fig:1DPaths_xh2_080} shows the results of the flame segment analysis for the high $\XHF$ case. An equivalent version of \cref{fig:segment_analysis} for this case is provided in \cref{supp_fig:segment_analysis} in the supplementary material. Similar to the low $\XHF$ case, the \ce{HNO} path is the major \ac{NO} producer, followed by the \ce{NH} path. Although the de\ac{NOx} path shows a decreased importance compared to the low $\XHF$ case, it still represents the major consumption path. Again, the \ac{1D} flame and the segment in the neutrally curved region as very similar. For the positively curved region, the production pathways are enhanced, while the consumption pathways remain at a comparable level as the \ac{1D} flame. For the negatively curved region, the relative importance of the de\ac{NOx} pathway increases, leading to an overall consumption of \ce{NO} at lower values of $C_{\ce{H2O}}$. However, due to the poor alignment of consumption and production regions, high production rates are observed at higher values of progress variable. Compared to the low $\XHF$ case, the difference between the total production rate in the positively and negatively curved regions is decreased by a factor of 4. On the macroscopic scale, this leads to the lower spread in \cref{fig:statistics} discussed before, and to the overall increase of the mean exhaust mass fraction of \ce{NO} shown in \cref{fig:1DComparison}.

\begin{figure}[h!]
	\centering
	\includegraphics[scale=1]{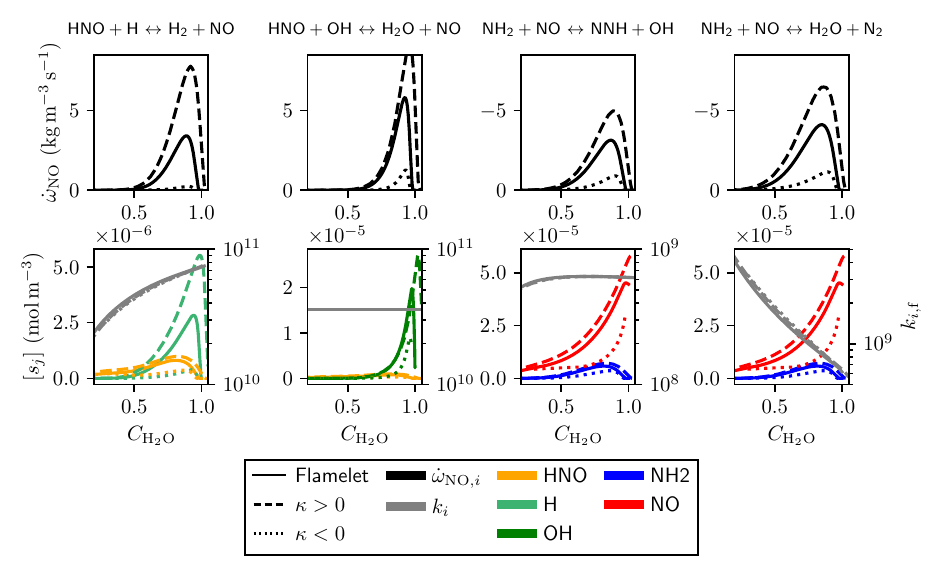}
	\caption{Production rate $\dot{\omega}_{\mathrm{NO}, i}$ (top row) and its decomposition into concentration $[s_j]$ (bottom, left axis) and forward reaction rate coefficient $k_{i, \rm f}$ (bottom, right axis) for the two dominant production and consumption reactions. Solid lines represent the value for the \ac{1D} reference case, dashed and dotted lines represent values for the positively and negatively curved regions, respectively. Conditions: $\XHF=0.4, \phi=0.6, \Tu=\SI{298}{\kelvin}, p=\SI{1}{\bar}$.}
	\label{fig:reac_anal}
\end{figure}

Finally, it remains unclear whether the changes in production rates are an effect of the reduced local temperature or of the changes in local concentrations. In general, the \ac{NO} production rate of the elementary reaction $i$ can be formulated as 
\begin{equation}
	\dot{\omega}_{\mathrm{NO}, i} = M_{\ce{NO}} \nu_{\ce{NO}_i} \left(k_{i,\rm f} \prod_{s_j \in \mathcal{S}_i} [s_j]^{\nu_{s_j}'} - k_{i,\rm b} \prod_{s_j \in \mathcal{S}_i} [s_j]^{\nu_{s_j}''} \right)\,,
\end{equation}	
where $M_{\ce{NO}}$ is the molar mass of \ce{NO}, $\nu_{\ce{NO}_i}$ is the global stoichiometric coefficient of \ac{NO} in reaction $i$, and $k_{i,\rm f}$ and $k_{i,\rm b}$ are the forward and backward rate coefficients of reaction $i$, respectively. Further, $s_j$ is a species from the set of species $\mathcal{S}_i$ considered in reaction $i$, with the concentration $[s_j]$ and the stoichiometric coefficients $\nu_{s_j}'$ and $\nu_{s_j}''$ in the reactants and products, respectively. Here, $k_{i,\rm f}$ and $k_{i,\rm b}$ are dependent on temperature. \Cref{fig:reac_anal} presents $\dot{\omega}_{\mathrm{NO}, i}$ (top row) for the two dominant production and the two dominant consumption reactions, and their decomposition into $[s_j]$ and $k_{i,\rm f}$ (bottom row) for the flamelet and the segments in positively and negatively curved regions. Note that only forward rates are shown, since the backward reactions are negligible. It becomes evident that the absolute decrease in production rates from positively to negatively curved regions is mostly induced by a change in concentration, since $k_i$ does not show significant changes for the difference segments. Especially the \ce{H} radical shows a significant decrease in concentration for $\kappa<0$ compared to $\kappa>0$, most likely related to its high diffusivity. 

%% file: chapters/04_Conclusions.tex
\section{Conclusions}

In the presented study, the influence of \acp{IFI} on \ac{NO} formation in laminar premixed \ac{NH3}/\ac{H2}/air flames under lean, ambient conditions ($\phi = 0.6, \Tu=\SI{298}{\kelvin}, p=\SI{1}{\bar}$) and varying hydrogen fractions in the fuel blend was analyzed. Across all cases, the influence of \acp{IFI} is visible, leading to local temperature overshoots and distinct variations in the \ac{NO} formation, which, however, are decreasing with increasing $\XHF$. Although the local mass fraction of \ac{NO} is increased by up to 49\% in the positively curved regions, the mean mass fraction at the outlet is 5\% below the exhaust gas values from a \ac{1D} unstretched flame. This is related to the strong decrease of \ac{NO} formation in the negatively curved regions. For the high $\XHF$ case, the peaks of \ac{NO} in the positively curved regions, but especially the troughs in the negatively curved regions are less pronounced. This leads to a small relative increase of the mean mass fraction in the post-flame region of the \ac{2D} flame compared to the \ac{1D} flame. 

Next, a flame segment analysis was conducted for the low and high $\XHF$ cases. The analysis reveals the formation pathway via \ce{HNO} as the most important, followed by the \ce{NH} pathway, regardless of $\XHF$ and the local curvature. Consumption is dominated by the de\ac{NOx} pathway followed by the \ce{N2O} pathway. Additionally to the absolute changes of \ac{NO} production rates depending on the local curvature, the relative importance of the de\ac{NOx} pathway increases in the negatively curved region. Furthermore, in the negatively curved region, production profiles shift to lower values of progress variable. For the low $\XHF$ case, this leads to an annihilation of production and consumption and hence, lower \ac{NO} mass fraction. For the high $\XHF$ case, this annihilation effect is less effective, leading to higher \ac{NO} mass fraction. Finally, the decrease of \ac{NO} production is found to be mainly driven by changes of the local radical concentrations, rather than changes of the temperature-dependent reaction rate coefficients.

%% file: chapters/05_Acknowledgements.tex
\section{Acknowledgement}
\label{sec:Acknowledgements}

TL, ND, MG, and HP gratefully acknowledge the received funding from the European Research Council (ERC) under the European Union’s Horizon 2020 research and innovation program (Grant agreement No. 101054894). TLH acknowledges the generous support by DFG (IRTG 2983 Hy-Potential: Hydrogen - Fundamentals of Production, Storage \& Transport, Applications, and Economy).

The authors gratefully acknowledge the computing time provided to them at the NHR Center NHR4CES at RWTH Aachen University (project numbers p0020340 and p0020410). This is funded by the Federal Ministry of Education and Research, and the state governments participating on the basis of the resolutions of the GWK for national high performance computing at universities (www.nhr-verein.de/unsere-partner).

%% file: chapters/99_Supplementary.tex
\section{Supplementary material}

\begin{figure}[h!]
	\centering
	\centering
	\begin{subfigure}[b]{0.49\textwidth}
		\centering
		\includegraphics[width=\textwidth]{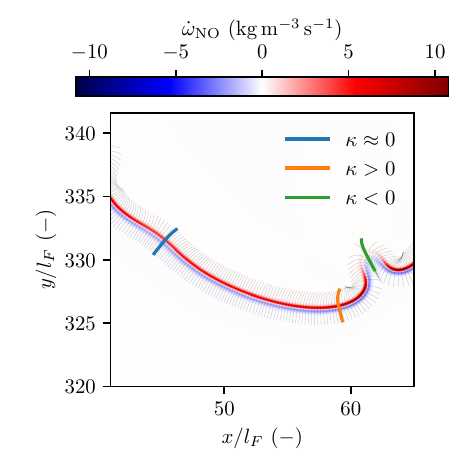}
		\caption{}
		\label{fig:selected_segments}
	\end{subfigure}
	\hfill
	\begin{subfigure}[b]{0.49\textwidth}
		\centering
		\includegraphics[width=\textwidth]{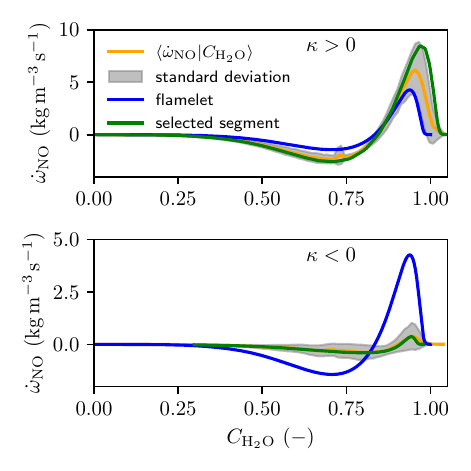}
		\caption{}
		\label{fig:segment_stats}
	\end{subfigure}
	\caption{Assessment of the statistical validity of the segment selection procedure. (a) Constructed segments (thin gray lines) on the spatial distribution of the \ac{NO} production term $\dot{\omega}_{\ce{NO}}$. The path density is reduced by a factor of 10 for visualization purposes. Thick colored lines depict the selected representative segments. (b) Mean of $\dot{\omega}_{\ce{NO}}$ conditioned on $C_{\ce{H2O}}$, its standard deviation within the category, the \ac{1D} reference flame, and the selected representative segment for the group of positive (top) and negative (bottom) curved segments. Note that the scales in (b) are different across the two categories to capture their individual details. Conditions: $\XHF=0.4, \phi=0.6, \Tu=\SI{298}{\kelvin}, p=\SI{1}{\bar}$.}
	\label{fig:segment_analysis}
\end{figure}

\begin{table}[h!]
	\centering
	\caption{NO reactions incorporated in this study, grouped by their associated pathways.}
	\begin{tabular}{crclc}
		\hline\noalign{\smallskip}
		\textbf{Reaction no.} & \multicolumn{3}{c}{\textbf{Chemical Formulation}} & \textbf{Pathway}\\
		\noalign{\smallskip}\hline\noalign{\smallskip}
		69 & \ce{NH2 + HNO} & $\leftrightharpoons$ & \ce{NH3 + NO} & \multirow{7}*{\ce{HNO}}\\
		173 & \ce{HNO} & $\leftrightharpoons$ & \ce{NO + H} & \\
		174 & \ce{HNO + O} & $\leftrightharpoons$ & \ce{NO + OH} & \\
		175 & \ce{HNO + H} & $\leftrightharpoons$ & \ce{NO + H2} & \\
		176 & \ce{HNO + OH} & $\leftrightharpoons$ & \ce{NO + H2O} & \\
		177 & \ce{HNO + O2} & $\leftrightharpoons$ & \ce{NO + HO2} & \\
		180 & \ce{HNO + NH} & $\leftrightharpoons$ & \ce{NH2 + NO} & \\
		\hline
		38 & \ce{NH + O} & $\leftrightharpoons$ & \ce{NO + H} & \multirow{2}*{\ce{NH}}\\
		43 & \ce{NH + O2} & $\leftrightharpoons$ & \ce{ NO + OH} & \\
		\hline
		48 & \ce{NH + NO} & $\leftrightharpoons$ & \ce{N2 + OH} & \multirow{3}*{de\ce{NO}$_x$}\\
		64 & \ce{NH2 + NO} & $\leftrightharpoons$ & \ce{NNH + OH} & \\
		65/66 & \ce{NH2 + NO} & $\leftrightharpoons$ & \ce{N2 + H2O} & \\
		\hline
		47 & \ce{NH + NO} & $\leftrightharpoons$ & \ce{N2O + H} & \multirow{2}*{\ce{N2O}}\\
		79 & \ce{N2H2 + NO} &  $\leftrightharpoons$ & \ce{N2O + NH2} & \\
		\hline
		34 & \ce{N + O2} & $\leftrightharpoons$ & \ce{NO + O} & \multirow{3}*{Zeldovich}\\
		35 & \ce{N + OH} & $\leftrightharpoons$ & \ce{NO + H} & \\
		36 & \ce{N + NO} & $\leftrightharpoons$ & \ce{N2 + O} & \\
		\noalign{\smallskip}\hline
	\end{tabular}
	\label{supp_tab:reactions}
\end{table}

\begin{figure}[h!]
	\centering
	\centering
	\begin{subfigure}[b]{0.49\textwidth}
		\centering
		\includegraphics[width=\textwidth]{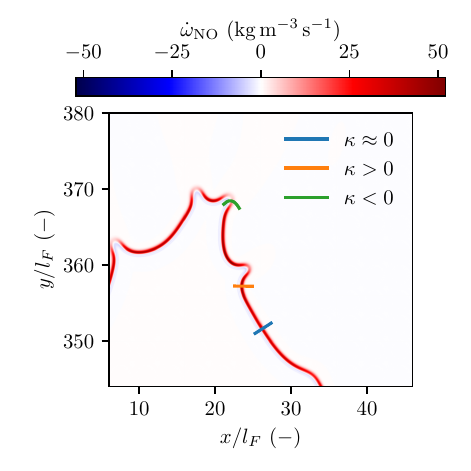}
		\caption{}
		\label{supp_fig:selected_segments}
	\end{subfigure}
	\hfill
	\begin{subfigure}[b]{0.49\textwidth}
		\centering
		\includegraphics[width=\textwidth]{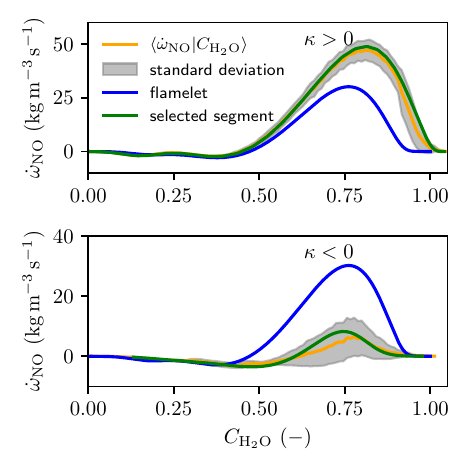}
		\caption{}
		\label{supp_fig:segment_stats}
	\end{subfigure}
	\caption{Assessment of the statistical validity of the segment selection procedure. (a) Selected representative segments on the spatial distribution of the \ac{NO} production term $\dot{\omega}_{\ce{NO}}$. (b) Mean of $\dot{\omega}_{\ce{NO}}$ conditioned on $C_{\ce{H2O}}$, its standard deviation within the category, the \ac{1D} reference flame, and the selected representative segment for the group of positive (top) and negative (bottom) curved segments. Note that the scales in (b) are different across the two categories to capture their individual details. Conditions: $\XHF=0.8, \phi=0.6, \Tu=\SI{298}{\kelvin}, p=\SI{1}{\bar}$.}
	\label{supp_fig:segment_analysis}
\end{figure}